\documentclass[12pt]{article}

\newenvironment{wileykeywords}{\textsf{Keywords:}\hspace{\stretch{1}}}{\hspace{\stretch{1}}\rule{1ex}{1ex}}

\usepackage{amsmath,amssymb}
\usepackage{mathtools}
\usepackage{graphicx}
\usepackage{color}
\usepackage{dcolumn}
\usepackage{bm}
\usepackage[numbers,super,comma,sort&compress]{natbib}
\usepackage[nolists, nomarkers, figuresfirst]{endfloat}
\definecolor{background-color}{gray}{0.98}

\makeatletter
\renewcommand*{\@fnsymbol}[1]{\ensuremath{\ifcase#1\or \dagger\or \ddagger\or *\or
   \mathsection\or \mathparagraph\or \|\or **\or \dagger\dagger
   \or \ddagger\ddagger \else\@ctrerr\fi}}
\makeatother

\title{Mode-Tracking Based Stationary-Point Optimization}
\author{Maike Bergeler\thanks{ETH Z\"urich, Laboratorium f\"ur Physikalische Chemie, Vladimir-Prelog-Weg 2, 8093 Z\"urich, Switzerland}, 
Carmen Herrmann\thanks{University of Hamburg, Institute of Inorganic and Applied Chemistry, Martin-Luther-King-Platz 6, 20146 Hamburg, Germany}
 \thanks{corresponding authors: carmen.herrmann@chemie.uni-hamburg.de, markus.reiher@phys.chem.ethz.ch}
 and Markus Reiher\footnotemark[1] \footnotemark[3]}

\begin{document}

\maketitle

\begin{abstract}
In this work we present a transition-state optimization protocol based on the Mode-Tracking algorithm 
[{\it J. Chem. Phys.}, {\bf 2003}, {\it 118}, 1634]. 
By calculating only the eigenvector of interest instead of diagonalizing the full Hessian matrix
and performing an eigenvector following search based on the selectively calculated vector, 
we can efficiently optimize transition-state structures.
The initial guess structures and eigenvectors are either chosen from 
a linear interpolation between the reactant and product structures, 
from a nudged-elastic band search, from a constrained-optimization scan, or
from the minimum-energy structures.
Alternatively, initial guess vectors based on chemical intuition may be defined.
We then iteratively refine the selected vectors by the Davidson subspace iteration technique. 
This procedure accelerates finding transition states for large molecules of a few hundred atoms.
It is also beneficial in cases where the starting structure is very different from the 
transition-state structure or
where the desired vector to follow is not the one with lowest eigenvalue. 
Explorative studies of reaction pathways are feasible by 
following manually constructed molecular distortions.
\end{abstract}

\begin{wileykeywords}
Transition-state search, Mode-Tracking, Davidson subspace iteration, Eigenvector following, Reaction kinetics.
\end{wileykeywords}

\clearpage


\begin{figure}[h]
\centering
\colorbox{background-color}{
\fbox{
\begin{minipage}{1.0\textwidth}
\includegraphics[width=110mm]{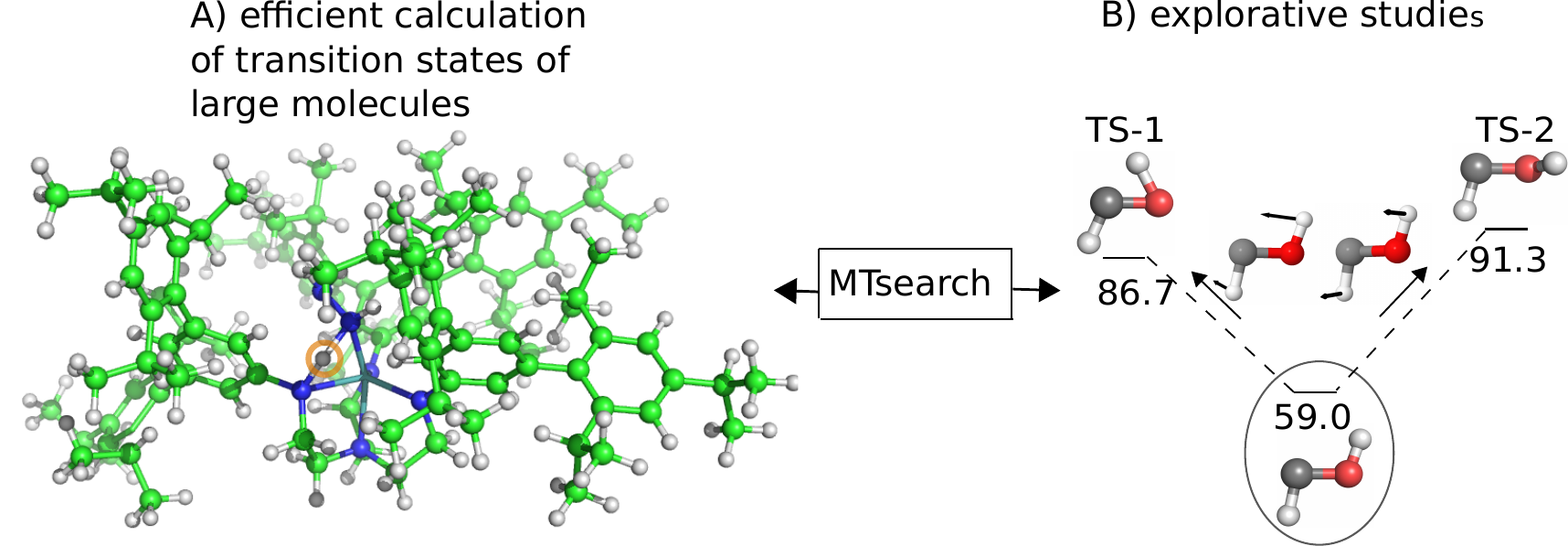}
\\
The access of specific normal modes out of all vibrational modes
opens up new possibilities for the search of stationary points. 
Since the calculation of the complete Hessian matrix is avoided, transition-state structures of large molecules can be obtained. 
Because of the possibility to start the transition-state search from one
initial guess structure, an explorative investigation of the potential energy surface is feasible.
\end{minipage}
}}
\end{figure}

  \makeatletter
  \renewcommand\@biblabel[1]{#1.}
  \makeatother

\bibliographystyle{apsrev}

\renewcommand{\baselinestretch}{1.5}
\normalsize

\clearpage

\section*{\sffamily \Large INTRODUCTION} 
The optimization of transition-state structures (TSs) is key to the understanding of mechanisms and kinetics of chemical reactions on a computational basis.
Transition states are defined as first-order saddle-point structures located on the minimum (reaction) energy path between reactants and products. 
First-order saddle points are characterized by one negative eigenvalue of the matrix of second partial derivatives of the electronic energy with respect
to the Cartesian nuclear coordinates, i.e. of the Hessian.
Reactants and products are local minima on the Born--Oppenheimer potential energy surface (PES).
The energy differences between a TS and two minima of an elementary reaction are the activation energy barriers. 
They should in principle be evaluated from the Gibbs free energy, but are approximated here, as in most quantum-chemical studies, by the electronic energy at zero
Kelvin (neglecting temperature and entropy contributions).

Numerous methods have been developed to efficiently find TSs. Examples are interpolation methods \cite{Halgren1977}, 
eigenvector following (EVF) \cite{Cerjan1981, Simons1983, Wales1992, Wales1993, Jensen1995}, string methods \cite{Henkelman2000}, and the scaled hypersphere search method \cite{Ohno2004}.
The existing TS search methods can be divided into those that start from one structure (often called single-ended methods) or those that require at least two starting structures, 
usually reactant and product structures (double-ended methods). 
Double-ended TS search algorithms are often based on interpolation methods such as linear (LST \cite{Halgren1977, Peng1993}) or quadratic synchronous transit (QST \cite{Halgren1977}),
 string methods or nudged elastic band (NEB \cite{Henkelman2000,Henkelman2000_2}) algorithms. 
Since the double-ended methods usually show slow convergence near a TS \cite{Jensen}, they are mainly employed to find a guess structure close to the TS, which 
is then refined by a more efficient single-ended method, such as EVF. 
Hence, it is beneficial to combine single-ended and double-ended methods for TS searches.

In most of the EVF-based methods, the full Hessian of the transition-state guess structure is calculated
to obtain the exact vibrational mode to follow. 
For large molecules, the complete Hessian calculation is computationally demanding as the calculation of the elements of the Hessian matrix is very time consuming
within a first-principles electronic-structure description.
Therefore, several algorithms have been developed to circumvent the calculation of the full Hessian in structure-optimization algorithms. 
A quasi-Newton--Raphson method has been introduced by Broyden.\cite{Broyden1967,Fletcher1980} In this method,
an approximate Hessian is built from gradients only and then updated (according to Bofill \cite{Bofill1994} and Powell \cite{Powell1971}) by 
the gradients of intermediate points obtained during the optimization. These methods reduce the computational effort significantly,
but for large molecules a further reduction of the computational cost is desirable.

Recently, Sharada et al. \cite{Sharada2012} introduced an approximate-Hessian approach
based on the tangent of the transition-state guess structure determined by an interpolation between reactant and product structures and by local curvature information.
This approximate Hessian approach combined with the growing string method turned out to be computationally less 
expensive than previous Hessian approximations.\cite{Zimmerman2013, Zimmerman2013_2} 

Since the efficiency of a TS search depends strongly on the initial Hessian,
a main goal is to set up an approximate Hessian matrix that resembles the exact one as closely as possible. 
In 2002, we proposed an algorithm based on Davidson subspace diagonalization for the selective calculation of eigenvectors of the mass-weighted Hessian
 based on predefined molecular distortions.\cite{Reiher2003}
This so-called Mode-Tracking scheme turned out to be very efficient in vibrational spectroscopy
\cite{Reiher2003, Neugebauer2004, Neugebauer2004_2, Reiher2004, Reiher2005, Adler2006, Herrmann2006_2, Herrmann2007, Herrmann2008}. 
Because of the straightforward and flexible implementation, Mode-Tracking was implemented in a semi-numerical fashion.\cite{Reiher2003}
At the same time, Deglmann and Furche \cite{Deglmann2002} presented an implementation of a fully analytical Davidson subspace diagonalization of the Hessian for
the optimization of its lowest eigenvalue required for the identification of stationary points.

Very recently, Sharada et al. \cite{Sharada2014} described a semi-numerical Davidson subspace iteration method to obtain 
selected information of the Hessian spectrum, which is identical to Mode-Tracking \cite{Sharada2014_2}.
For transition-state optimizations, Sharada et al.\cite{Sharada2014} extract the guess mode from the coordinates along the pathway obtained from the freezing-string method (FSM).
In contrast to the Hessian approach presented by Sharada et al. \cite{Sharada2014}, we here develop a Mode-Tracking-based TS and
minimum localization algorithm that can iteratively refine a specific eigenvector of interest, which does not have to be the one with lowest eigenvalue.
Our algorithm can be executed in an explorative fashion as we can circumvent the NEB or FSM calculation by starting from only one minimum-energy structure and by
following several eigenvectors in one optimization in parallel. We will demonstrate these capabilities at the example of the isomerization reactions of formaldehyde, which 
has been studied as a benchmark system for automated transition-state search algorithms \cite{Davis1990, Bondensgard1996, Quapp1998, Maeda2013_2}.
Subsequently, we investigate an internal proton-transfer reaction in a hydrazine complex, which is an intermediate in the Schrock N$_2$-fixation 
catalytic cycle.\cite{Yandulov2002,Yandulov2003,Schrock2005}
At this example, we examine the applicability of our algorithm for finding TSs for large molecules 
(the Schrock catalyst contains 284 atoms). Although smaller model complexes can be generated, the smallest ones, which 
resemble the structure of the original catalyst, still comprise 41 atoms.

We choose these examples to highlight the capabilities of the Mode-Tracking-based approach to TS searches, which improves on existing methods rather than proposing a new TS search algorithm.
Hence, validating the performance of our Mode-Tracking version of existing TS search algorithms at standard TS test sets \cite{Baker1996, Zhao2005, Guner2003, Zimmerman2013_3} 
is neither needed nor necessary.

This paper is organized as follows: In the next section, the Mode-Tracking algorithm and the theoretical background of transition-state optimizations are described. 
After the subsequent Computational Methodology section, results are reported for our benchmark reactions.

\section*{\sffamily \Large THEORY}

The main idea of the algorithm to be described is to find transition-state structures by following only 
certain eigenvectors of the Hessian matrix selectively calculated by Mode-Tracking.
For stationary structures, the harmonic vibrational frequencies and the corresponding eigenvectors of a system can be obtained by solving the following eigensystem,

\begin{equation}
\mathbf{H}\mathbf{Q}_k=\lambda_k\mathbf{Q}_k,
\label{0}
\end{equation}
where $\mathbf{H}$ is the mass-weighted Cartesian Hessian, $\lambda_k$ are the eigenvalues,
and the eigenvectors $\mathbf{Q}_k$ are the mass-weighted vibrational normal modes.
For non-stationary structures, for which the length of the geometry gradient is nonzero, Eq.\ (\ref{0}) cannot be related to the vibrational properties of a molecule, but the
eigenpairs (eigenvalues and eigenvectors) of $\mathbf{H}$ still characterize the PES.

In Mode-Tracking, the eigenpairs of interest are obtained through a Davidson-type subspace iteration method\cite{Reiher2003, Reiher2004},
in which the original Hessian matrix $\mathbf{H}$ is transformed to the (reduced-dimensional) 
Davidson matrix $\mathbf{\tilde{H}}^{i}$,

\begin{equation}
\mathbf{\tilde{H}}^{i}=(\mathbf{B}^i)^T\mathbf{H}\mathbf{B}^i\equiv\displaystyle(\mathbf{B}^i)^T\mathbf{\Sigma}^{i},
\end{equation}
where $i$ denotes the $i$-th iteration step. $\mathbf{B}^i$ is a matrix whose columns contain collective displacement vectors $\mathbf{b}^l$ ($l$\,=\,1,...,$i$) 
along the $3M$ (mass-weighted) nuclear Cartesian basis vectors ($M$ is the number of atoms). 
In our semi-numerical implementation, $\mathbf{\Sigma}^i$ contains all vectors $\boldsymbol{\sigma}^l$, which collect
the (numerical) derivatives of the (analytical) Cartesian gradient $\mathbf{g}$ of the total electronic energy with respect to the corresponding 
collective displacement vector $\mathbf{b}^l$,
\begin{equation}
\boldsymbol{\sigma}^l=\mathbf{H}\mathbf{b}^l = \displaystyle{\begin{pmatrix} \displaystyle\sum_n H_{1,n} b_n^l \\ \displaystyle\sum_n H_{2,n} b_n^l \\ ... \\
\displaystyle \sum_n H_{3M,n} b_n^l \end{pmatrix}} \\
= \begin{pmatrix} \frac{\displaystyle\partial}{\displaystyle\partial \mathbf{b}^l} \frac{\displaystyle\partial E_{\mathrm{el}}}{\displaystyle\partial R_1}\\ 
\frac{\displaystyle\partial}{\displaystyle\partial \mathbf{b}^l} \frac{\displaystyle\partial E_{\mathrm{el}}}{\displaystyle\partial R_2} 
\\ ... \\ \frac{\displaystyle\partial}{\displaystyle\partial \mathbf{b}^l} \frac{\displaystyle\partial E_{\mathrm{el}}}{\displaystyle\partial R_{3M}} \end{pmatrix}\\
= \frac{\partial}{\partial\mathbf{b}^l}\mathbf{g}.
\end{equation}

By solving
\begin{equation}
\mathbf{\tilde{H}}^{i} \mathbf{c}_{k}^i = \lambda_{k}^i\mathbf{c}_{k}^i,
\end{equation}
for the eigenvectors $\mathbf{c}_{k}^i$ and eigenvalues $\lambda_{k}^i$.
In the $i$-th iteration step, Mode-Tracking calculates the approximate $\it{k}$-th normal mode $\mathbf{Q}_k^i$ as
\begin{equation}
\mathbf{Q}_{k}^{i}=\sum_{l=1}^ic_{k,l}^{i}\mathbf{b}^l.
\label{eq:2}
\end{equation}

New basis vectors $\mathbf{b}^{i+1}$ are generated from the residuum vector, 
\begin{equation}
\mathbf{r}_{k}^i=[\mathbf{\tilde{H}}^{i}-\lambda_{k}^{i}]\mathbf{Q}_{k}^i,
\end{equation}
after applying a preconditioner $\mathbf{X}^i$ to it\cite{Reiher2004},
\begin{equation}
\mathbf{b}^{i+1}=\mathbf{X}^i\mathbf{r}_k^i.
\end{equation}

The initial guess mode $\mathbf{b}^1$ can be obtained from the LST, which linearly interpolates between the reactant and product structures, or
from other path methods such as NEB.
Let $\mathbf{R}^{\textrm{(nmw)}}_j$ be the non-mass-weighted '(nmw)'
Cartesian coordinates of a structure $j$ on the PES, 
then the initial normalized, non-mass-weighted mode is 
constructed from the coordinate differences between this 
point and each of its neighboring points, 
$\mathbf{R}^{\textrm{(nmw)}}_{j+1}=\{{R}^{\textrm{(nmw)}}_{k,j+1} \}$ and 
$\mathbf{R}^{\textrm{(nmw)}}_{j-1}=\{{R}^{\textrm{(nmw)}}_{k,j-1}\}$, 

\begin{equation}
\mathbf{b}_{j}^{\textrm{(nmw)},1} ={1\over 2}\left[\frac{(\mathbf{R}^{\textrm{(nmw)}}_{j+1}-\mathbf{R}^{\textrm{(nmw)}}_{j})}{|\mathbf{R}^{\textrm{(nmw)}}_{j+1}-\mathbf{R}^{\textrm{(nmw)}}_{j}|}+\frac{(\mathbf{R}^{\textrm{(nmw)}}_{j}-\mathbf{R}^{\textrm{(nmw)}}_{j-1})}{|\mathbf{R}^{\textrm{(nmw)}}_{j}-\mathbf{R}^{\textrm{(nmw)}}_{j-1}|}\right]=\left\{b_{k,j}^{\textrm{(nmw)}} \right\},
~\quad k=1,\dots,3M.
\end{equation}
This mode is then mass-weighted,
\begin{equation}
\mathbf{b}^1_{j} = \left\{ \frac{b_{k,j}^{\textrm{(nmw)}}{\sqrt{m_k}}}{\sum_{k=1}^{3M}\left(b_{k,j}^{\textrm{(nmw)}}{\sqrt{m_k}}\right)^2} \right\}, ~\quad k=1,\dots,3M,
\end{equation}
where $m_k$ is the mass of the $k$-th atomic nucleus.

In general, Mode-Tracking can either optimize the mode with largest overlap with the initial guess vector or 
the one with largest overlap with the approximate eigenvector chosen in the last iteration (root-homing).
If the initial guess vector differs strongly from the normal mode of the transition-state structure, the second option might be more suited to find a TS.

The eigenvector following algorithm \cite{Jensen1995} is then employed to steer the optimization into the direction
of the TS and to finally locate it. Newton--Raphson steps along the converged Mode-Tracking eigenvector, which is referred to as transition vector, are carried out to maximize the energy 
in this direction, while in all directions orthogonal to the transition direction the structure is relaxed \cite{Wales1994}.
For this, we project out the gradient along the transition vector, $\mathbf{g}_{\mathrm{TS}}$, from the total molecular gradient, 

\begin{equation}
 \mathbf{g}=\{g_k\}=\left\{\frac{\partial E}{\partial R_k}\right\}, \qquad  k= 1,...,3M. 
 \label{eq:9}
\end{equation}

To obtain the components of the molecular gradient that are orthogonal to the eigenvector, $\mathbf{g}_{\mathrm{ort}}$, 
we subtract the gradient part along the transition vector from the original molecular gradient and obtain

\begin{equation}
 \mathbf{g}_{\mathrm{ort}}^{({\rm nmw})}
=\mathbf{g}^{({\rm nmw})}-\underbrace{\mathbf{Q}^{({\rm nmw})}_{\mathrm{TS}}\mathbf{Q}_{\mathrm{TS}}^{({\rm nmw}),T}\mathbf{g}^{({\rm nmw})}}_{\mathbf{g}^{({\rm nmw})}_{\mathrm{TS}}}= (1-\mathbf{Q}^{({\rm nmw})}_{\mathrm{TS}}\mathbf{Q}_{\mathrm{TS}}^{({\rm nmw}),T})\mathbf{g}^{({\rm nmw})},
\end{equation}
where $\mathbf{Q}^{({\rm nmw})}_{\mathrm{TS}}$ is the selected eigenvector calculated with Mode-Tracking, which approximately points into the direction of the TS. This is done in no-mass-weighted coordinates.
The corresponding eigenvalue is $\lambda_{\mathrm{TS}}$.

Let $\mathbf{R}_0$ be the coordinates of the targeted stationary point, for which $\mathbf{g}_0\equiv\{(\partial E/\partial R_k)_{R_k=R_{0,k}} \}$ vanishes component-wise,
and $\mathbf{H}_0\equiv\{(\partial^2 E/\partial R_k\partial R_l)_{R_k=R_{0,k},R_l=R_{0,l}} \}$.
From a truncated Taylor series expansion of the potential energy around $E_0=E(\mathbf{R}_0)$ on the PES,
\begin{equation}
E(\mathbf{R})= E_0 + \mathbf{g}_0^T\boldsymbol{\Delta}\mathbf{R} +\frac{1}{2}\boldsymbol{\Delta}\mathbf{R}^T\mathbf{H}_0\boldsymbol{\Delta}\mathbf{R} + O(\boldsymbol{\Delta}\mathbf{R}^3).
\end{equation}
the coordinate displacement $\boldsymbol{\Delta}\mathbf{R}\equiv\mathbf{R}-\mathbf{R}_0$ that leads to a stationary point ($dE(\mathbf{R})$/$d\mathbf{R}=\mathbf{0}$), 
\begin{equation}
{\boldsymbol{\Delta}}\mathbf{R} = -\frac{\mathbf{g}_0}{\mathbf{H}_0},
\end{equation}
can be derived. $\mathbf{R}_0$ is the position of a stationary structure, $\mathbf{g}_0$ its gradient and $\mathbf{H}_0$ its Hessian.
$\boldsymbol{\Delta}\mathbf{R}$ can be split into a direction parallel to the transition vector, $\boldsymbol{\Delta}{\bf R}_{\mathrm{TS}}$, 
and into all other directions. The step in the direction of the transition vector reads
\begin{equation}
\boldsymbol{\Delta}{\bf R}_{\mathrm{TS}}=-\frac{\mathbf{g}_{\mathrm{TS}}}{\lambda_{\mathrm{TS}}}.
\end{equation}
The energy in direction of the selected mode is maximized if $\lambda_{\mathrm{TS}}$ is negative.
If we do not start the EVF procedure from a structure close to the TS, but, for instance, from a minimum structure, we must ensure that the transition vector is still followed uphill.
This can either be accomplished by employing the absolute value of $\lambda_{\mathrm{TS}}$
\begin{equation}
 \boldsymbol{\Delta}{\bf R}_{\mathrm{TS}}=\frac{\mathbf{g}_{\mathrm{TS}}}{|\lambda_{\mathrm{TS}}|},
\end{equation}
 or by employing Eq.\ (\ref{eq:1}) described below.

To improve on the convergence of the EVF optimization, Wales \cite{Wales1994} defined a Lagrangian with Lagrangian multipliers $\kappa_k$ for each degree of freedom $l$:
\begin{equation}
 L=-E_0-\mathrm{\bf g}_0^T{\bf \Delta R}-\frac{1}{2}{\bf \Delta R}^T{\bf H}_0{\bf \Delta R}+\frac{1}{2}\sum_{l=1}^{3M} \kappa_l(\Delta R_l^2-c_l^2).
\end{equation}

Wales' method employs the rational function by Banerjee\cite{Simons1983, Neal1984, Banerjee1985}, in which the Lagrangian 
multipliers are defined by the eigenvalues $\lambda_k$ and the gradient components $\mathbf{g}_k$ along the
eigenvectors,
\begin{equation}
 \kappa_l=\frac{1}{2}\displaystyle{\left(\lambda_l\pm\sqrt{\lambda_l^2+4\mathbf{g}_l^2}\right)}. 
\end{equation}

It appears to be more efficient \cite{Wales1991} to modify the equation to the following one:

\begin{equation}
 \kappa_k=\lambda_l\pm\frac{1}{2}|\lambda_l|\left(1+\sqrt{1+\frac{\displaystyle{4\mathbf{g}_l^2}}{\displaystyle{\lambda_l^2}}}\right)
\end{equation}
where '$+$' is for maximization and '$-$' for minimization.

Wales arrived at the following equation that describes the steps to be made along all degrees of freedom $l$,
\begin{equation}
 \mathbf{\Delta R}_l=\frac{\pm2\mathbf{g}_l}{|\lambda_l|\left(1+\sqrt{1+\frac{\displaystyle{4\mathbf{g}_l^2}}{\displaystyle{\lambda_l^2}}}\right)},
 \label{eq:1}
\end{equation}

where '$+$' leads to an uphill and '$-$' to a downhill energy step.
For a TS search, an uphill step along the desired mode (i.e., the approximate transition vector) is required,

\begin{equation}
 \mathbf{\Delta R}_{\mathrm{TS}}=\frac{+2\mathbf{g}_\mathrm{TS}}{|\lambda_{\mathrm{TS}}|\left(1+\sqrt{1+\frac{\displaystyle{4\mathbf{g}_\mathrm{TS}^2}}{\displaystyle{\lambda_{\mathrm{TS}}^2}}}\right)}.
 \label{eq:1}
\end{equation}

\section*{\sffamily \Large Computational Methodology}

\subsection*{\sffamily The {\sc MTsearch} program}

\begin{figure}[H]
\begin{center}
  \includegraphics[scale=0.8]{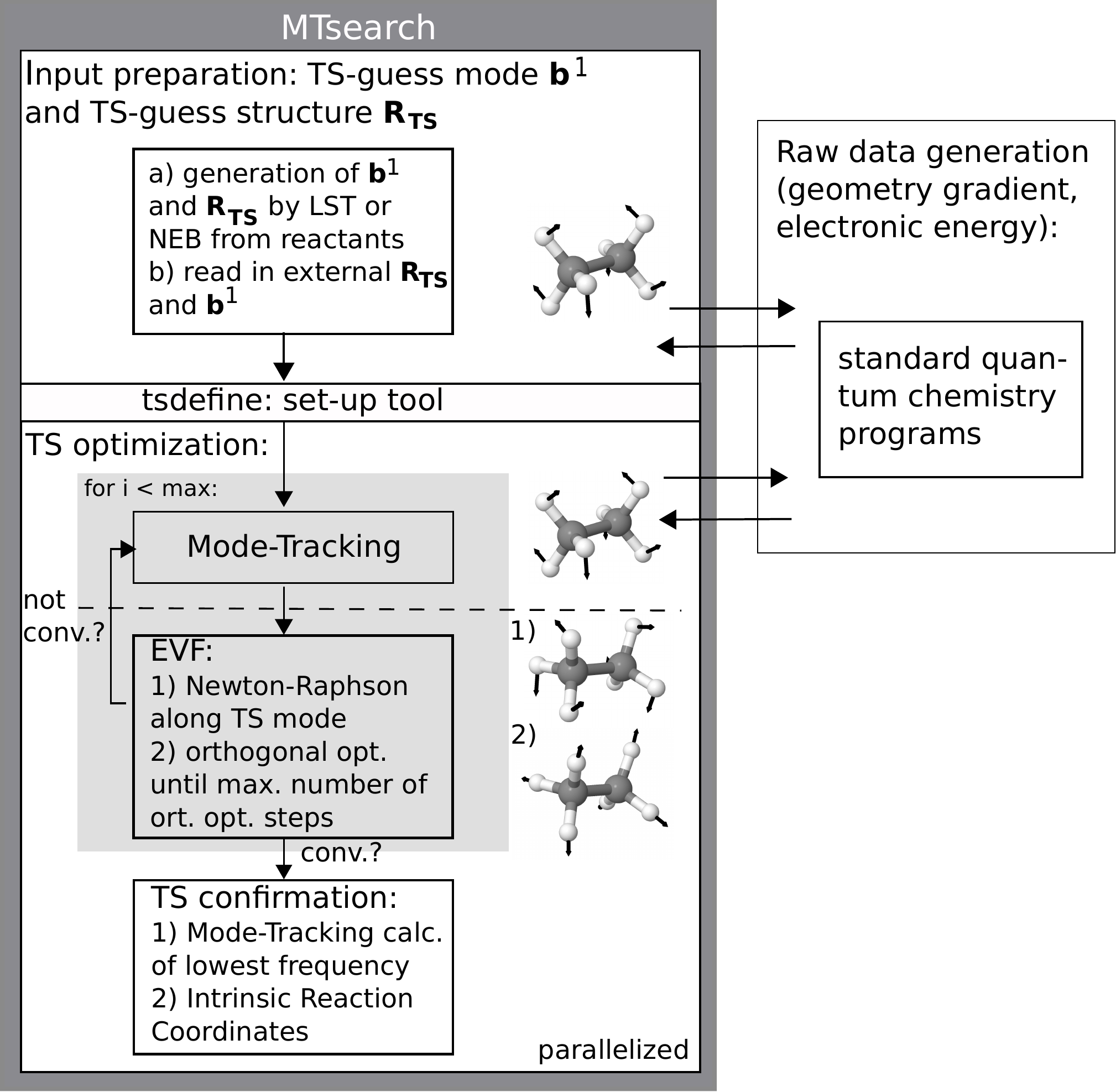} 
  \caption{Overview of the {\sc MTsearch} meta-program structure.}
\label{overview}
\end{center}
\end{figure}

We implemented the theory presented in the previous section in a computer program called {\sc MTsearch}. 
The program is based on the original Mode-Tracking program \cite{Reiher2003, Reiher2004,AKIRA},
which is currently available in its latest release as part of the {\sc MoViPac} package\cite{MOVIPAC}.
{\sc MTsearch} is a parallelized meta-program that accesses standard quantum-chemical programs for the calculation of gradients and electronic energies. 
The computational methodology for the generation of these raw data is described in detail in the next subsection. 
The algorithmic structure of {\sc MTsearch} is sketched in Figure \ref{overview}.

\begin{center}
[Figure 1 about here.]
\end{center}

A set-up tool, called {\sc tsdefine}, creates the necessary input files for an {\sc MTsearch} calculation. 
With {\sc tsdefine} we read in initial guess structures and, if available, initial modes.
The initial guess modes and structures can either be created within {\sc MTsearch}, from a LST or an NEB path, or 
by an external program, which provides guess structures and modes, e.g., based on a constrained optimization scan. 
The LST or NEB path consists of six to twelve nodes (that is, molecular structures on an (approximate) reaction path, including reactants and products), 
which we found to be a reasonable number. 
The spring forces in an NEB calculation are set to 0.02 a.u., and such a calculation is considered converged when
the difference between the gradient norm of the actual iteration and the previous one drops below 1$\times$10$^{-3}$ a.u. 

The first step of the TS optimization procedure is the Mode-Tracking optimization of the initial guess mode
to produce the corresponding minimal mode.
Mode-Tracking is assumed to have converged when the maximum element of the residuum vector 
drops below 5$\times$10$^{-3}$ a.u. and the change in the length of the residuum vector
drops below 5$\times$10$^{-6}$ a.u. One may also choose the convergence criteria corresponding to the last-added basis vector contribution
to the selected eigenvector or to the change in the eigenvalue between the last iterations.

After the calculation of a specific mode with Mode-Tracking, an EVF step is performed based on this converged mode.
For the Newton--Raphson step along the transition vector, $\boldsymbol{\Delta}\mathbf{R}_{\mathrm{TS}}$, we define a maximum step size of 0.2 {\AA}/$\sqrt{\rm{amu}}$, which is decreased to 0.1 {\AA}/$\sqrt{\rm{amu}}$
when the norm of $\mathbf{g}_{\mathrm{TS}}$ drops below 3$\times$10$^{-2}$ a.u., and to 0.05 {\AA}/$\sqrt{\rm{amu}}$ when the norm of $\mathbf{g}_{\mathrm{TS}}$ drops below 1$\times$10$^{-2}$ a.u.
For TS searches starting from minimum-energy structures, the first four Newton--Raphson steps are set to a maximum length of 1.0 {\AA}/$\sqrt{\rm{amu}}$, whenever the Hessian eigenvalue is positive or 
close to zero, i.e., no imaginary frequency with a large absolute magnitude is obtained.

After each Newton--Raphson step a predefined number of optimization steps orthogonal to the eigenvector is performed.
As default, a maximum of three iterations is chosen, if not otherwise mentioned.

If the norm of the total gradient is still above the threshold (default is 1$\times$10$^{-3}$ a.u.) after the predefined number of orthogonal optimization steps, 
another Mode-Tracking calculation is launched, for which the last converged Mode-Tracking eigenvector is chosen as default guess vector.
By default, a root-homing scheme selects the eigenvectors during the Mode-Tracking calculation with respect to the largest overlap with the initial one.
For comparison, we also employed a root-homing scheme in which the eigenvector is always compared to the previous one.

It would also be possible to reuse the same eigenvector for a predefined number of EVF steps, and/or to perform more than one EVF step between orthogonal optimizations.
This has not been explored in this study.
It is also possible to supply more information about the transition path direction to {\sc MTsearch} than only the first eigenvector 
(e.g., a sequence of structures which can for example easily be generated by a
haptic device \cite{Marti2009,Haag2014,Haag2014_2}). The guess vectors for the first few Mode-Tracking calculations are then chosen according to the predefined sequence of structures.
It has to be specified how many times the initial guess structure path shall be taken as reference for creating a guess mode, which is then refined by Mode-Tracking.  

The structures and normal modes were visualized with {\sc Pymol} \cite{Pymol} and {\sc Jmol} \cite{Jmol}, respectively. 

\subsection*{\sffamily Raw data generation}

All energies and gradients which are read as raw data by {\sc MTsearch} were calculated with density functional theory employing the 
{\sc Turbomole} program package (version 6.3.1)\cite{Ahlrichs1989}
with Ahlrichs' def2-SV(P), def2-SVP and def2-TZVP basis sets \cite{Weigend2005}. 
{\sc MTsearch} launches these calculations by system calls.
Restricted and unrestricted BP86 \cite{Becke1988,Perdew1986} all-electron Kohn--Sham calculations in combination with the resolution-of-the-identity technique were carried out.  
Self-consistent-field single-point calculations are considered to be converged when the total electronic energy difference between two
iteration steps is less then 10$^{-7}$ Hartree, if not otherwise indicated.
Molecular structure minimizations are considered converged when the norm of the geometry gradient is below 10$^{-4}$ a.u.
For the optimization of transition-state structures a geometry-gradient threshold of 10$^{-3}$ a.u. is chosen.

\subsection*{\sffamily Reference calculations}

For comparison, we performed {\sc Turbomole} (version 6.3.1)\cite{Ahlrichs1989} EVF calculations for comparison with the {\sc MTsearch} results.
Starting structures were chosen from the LST, NEB, or constrained optimization paths.
We carry out a single-point calculation on the starting structure and continue with a calculation of all vibrational modes with {\sc Turbomole}.
Then, we employ the trust-radius imaging method (the maximum radius and the trust radius are chosen between 0.1 {\AA}/$\sqrt{\rm{amu}}$ and 0.2 {\AA}/$\sqrt{\rm{amu}}$) 
to follow the lowest eigenvalue.  
We refer to this procedure in the following as ``standard EVF method''.
In these {\sc Turbomole} calculations, the BP86 \cite{Becke1988,Perdew1986} density functional is chosen with Ahlrichs' def2-SV(P), def2-SVP and def2-TZVP basis sets \cite{Weigend2005}.

Furthermore, we performed constrained optimizations by employing the {\sc Gaussian} \cite{gaussian09} program (version 09, Revision C.1) 
to obtain transition-state guess structures. Essentially, one internal coordinate was kept fixed at defined values and for all other degrees of freedom a 
structure optimization was carried out. Furthermore, intrinsic reaction coordinates were calculated with {\sc Gaussian}.
In these calculations we employed BP86 with the def2-SVP basis set.\cite{Schaefer1992,Schaefer1994}
We have chosen the default convergence criteria (scfconv=tight, which means that the energy difference between two
SCF iterations was less than 10$^{-8}$ Hartree, and that the structure optimizations were considered converged when the root-mean-square force acting 
on all atoms was below 3$\times$10$^{-4}$ a.u.).

We should note that we provide data for the eigenvalues of the Hessian as 'frequencies'
(reported in units of wave numbers). I.e., we take the square root of the eigenvalues, which corresponds
to a harmonic vibrational frequency for a stationary structure, even for {\it non}-stationary structures and denote it a 'frequency' for the sake of convenience
(eventually, these data become harmonic frequencies upon convergence of the stationary-structure optimization). Moreover, to highlight imaginary frequencies,
we add a minus sign in front of them (this is possible as the square of such a frequency still yields the correct eigenvalue of the Hessian matrix). 
Note also that we use the term 'mode' to denote an eigenvector of the Hessian matrix.

\section*{\sffamily \Large RESULTS}

\begin{figure}[H]
\begin{center}
  \includegraphics[scale=1.1]{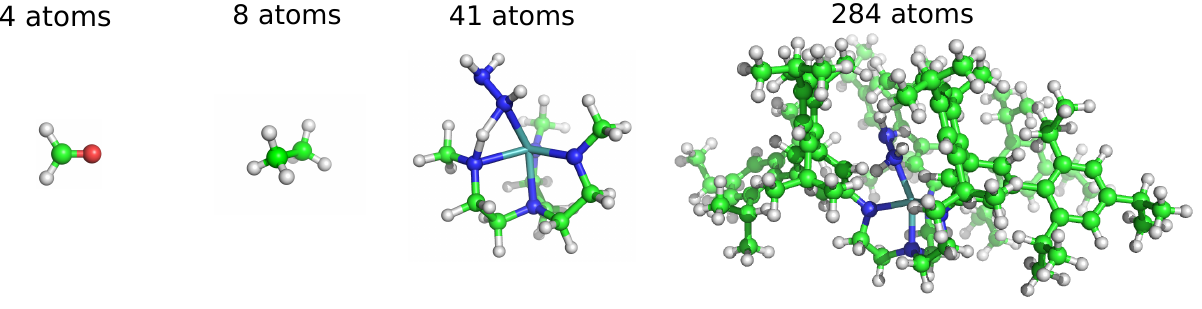} 
  \caption{Molecular models considered in this work: H$_2$CO (left), ethane (second from left), a small (third from left) 
and the full Schrock hydrazine tris(amido)amine Mo complex (right). Element color code: green, C; red, O; blue, N; cyan, Mo; white, H.}
\label{test-systems}
\end{center}
\end{figure}

To study the capabilities of {\sc MTsearch}, we have chosen four intramolecular reactions involving molecules of different sizes (4 atoms, 8 atoms, 41 atoms, and 
284 atoms; shown in Figure \ref{test-systems}).

\begin{center}
[Figure 2 about here.]
\end{center}

We start with the investigation of the rotational barrier in the C$_2$H$_6$ molecule, because the transition-state structure is well defined and the system is 
small, which allows us to investigate the suitable settings and thresholds for {\sc MTsearch}.
Next, we analyze the possibility of {\sc MTsearch} to optimize several transition-state structures starting from one minimum-energy structure using
hydroxymethylene as an example.

The last two reactions considered are possible side reactions of the Chatt--Schrock cycle of N$_2$ fixation at a molybdenum containing catalyst\cite{Yandulov2002,Yandulov2003},
in which N$_2$ is reduced to ammonia under acidic and reductive conditions.
Under these reaction conditions, it is possible that several unwanted intermediates are formed.
Exemplarily, we have chosen one possible side-reaction pathway, where one proton of N$_2$H$_4$ coordinated to molybdenum shifts to one of the amido nitrogens. 

The Schrock catalyst is ligated by a tetradentate hiptN$_3$N ligand (hipt\,= hexa-{\it iso}-propyl terphenyl). 
It has been intensively studied both experimentally \cite{Yandulov2003, Schrock2005, LeGuennic2005} and theoretically 
\cite{Studt2005, Reiher2005, LeGuennic2005, Schenk2008, Schenk2009, Schenk2009-2}.
Because of the relatively large system size of the hiptN$_3$N ligated hydrazine molybdenum complex 
(278 atoms), several smaller generic model system of the catalyst, in which the aryl substituents have been substituted, e.g.,
 by H atoms or CH$_3$ groups, have been studied.\cite{Schenk2009, LeGuennic2005, Khoroshun2002,Cao2005, Studt2005, Studt2005_2, Magistrato2007}
In the following, the small and large Schrock catalyst refer to the 
MeNCH$_2$CH$_3$N or hipt ligated molybdenum catalyst, respectively, with a hydrazine ligand as shown in Figure \ref{test-systems}.
Our focus is the optimization of transition-state structures for the proton-transfer reaction in the small and large Schrock catalysts. 

\subsection*{\sffamily \large Benchmark Example: C$_2$H$_6$ rotation}

We first calculated the transition-state structure of ethane rotation from staggered conformation to eclipsed conformation and back to staggered conformation.
To obtain starting structures and initial guesses for the Mode-Tracking scheme, we performed a linear synchronous transit in internal coordinates with
six nodes on the path including reactant and product structures (both staggered). 
Since only one dihedral angle is changed during the transition from one minimum structure to the other, the choice of internal coordinates
is very useful in this example. 
Due to the symmetry in the LST path, only three of the six structures are different.
The first (minimum), second, and third structures of the LST pathway were chosen as starting structures for EVF procedures
performed with {\sc MTsearch} and, for comparison, {\sc Turbomole}. 

Although the second structure of the LST pathway does not feature negative eigenvalues of the Hessian, the EVF algorithm optimization started from these structures 
converged towards the transition-state structure with both {\sc Turbomole} and {\sc MTsearch} (see Supporting Information for details).

If we start from the energy minimum structure, EVF relying on one negative eigenvalue is not able to find the TS, because the structure is far away from the quadratic region 
around the TS. Therefore, one would usually start from a guess structure closer to the TS.
By contrast, {\sc MTsearch} locates the TS starting from the energy-minimum structure with a mode specified by the LST and by a manually chosen mode corresponding to the rotation of one CH$_3$ group around the C$-$C axis. 
The initial-guess structures and converged TSs can be found in the Supporting Information.

We analyzed the effect of various parameters on the convergence of {\sc MTsearch}.
First, we investigated the optimal length of the first Newton--Raphson step.
If the starting structure is close to the energy minimum structure, the optimizer has to accomplish a larger step out of the minimum.
We observed that a first Newton--Raphson step size of 1.0--1.5 {\AA}/$\sqrt{\rm{amu}}$ is appropriate (see Supporting Information, Section 2, for details).

Next, we adapted the number of orthogonal optimization steps performed until the next mode is optimized by Mode-Tracking to
values between 2 and 10. 
To generalize the algorithm, we defined a protocol which stops the orthogonal optimization if the norm of the gradient for the optimization orthogonal to the transition path drops below 1$\times$10$^{-3}$ a.u., 
which means that the maximum number of orthogonal-optimization steps needs 
not to be reached. Then, the next Mode-Tracking calculation and 
Newton--Raphson step in the direction of the converged transition vector is performed.

\subsection*{\sffamily \large Explorative Example: Isomerization of H$_2$CO}

\begin{figure}[H]
\begin{center}
 \includegraphics[scale=0.8]{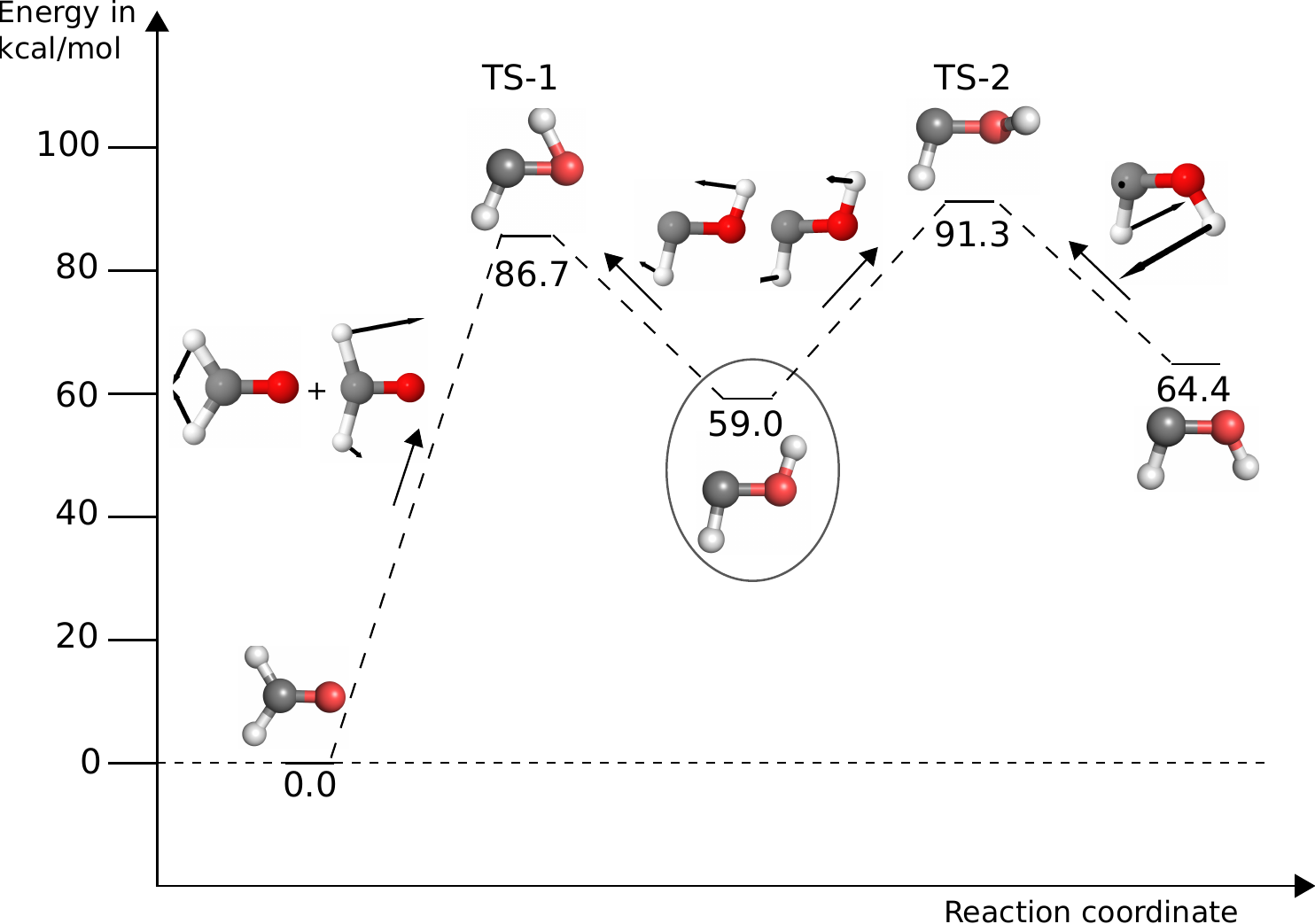}
 \caption{BP86/RI/def2-SV(P) reaction path from formaldehyde to trans-hydroxymethylene and cis-hydroxymethylene with transition-state structures optimized with {\sc MTsearch}.
The modes taken from a full vibrational analysis that lead to the TSs, {\bf TS-1} and {\bf TS-2}, are also depicted.
A maximum number of three orthogonal optimization steps has been chosen in {\sc MTsearch}. Element color code: gray, C; red, O; white, H.}
\label{path_h2co}
\end{center}
\end{figure}

In this section, we study the possibility to find several TSs with {\sc MTsearch} starting from one minimum-energy structure only.
We have chosen the isomerization reaction of formaldehyde to hydroxymethylene and a subsequent trans-/cis-isomerization of hydroxymethylene as an example (see Figure \ref{path_h2co}). 
The transition-state structures are well known \cite{Davis1990, Deng1993, Bondensgard1996, Quapp1998}.
Since two reaction pathways are possible from trans-hydroxymethylene, a selective way of choosing the eigenvector of interest is important.

\begin{center}
[Figure 3 about here.]
\end{center}

The trans-hydroxymethylene structure, from which we start the explorative TS search, 
can either undergo an internal hydrogen transfer from the oxygen atom to the carbon atom (over {\bf TS-1}) that leads to
formaldehyde or a rotation around the C-O axis that leads to cis-hydroxymethylene (over {\bf TS-2}). 
{\sc MTsearch} is able to locate both transition-state structures by following the modes
which are shown in Figure \ref{path_h2co} next to the arrows indicating the reaction direction.
The three lowest modes of the starting structure (obtained by a full Hessian calculation) have the following frequencies: 1100 cm$^{-1}$, 1188 cm$^{-1}$,
and 1317 cm$^{-1}$. The first one leads to {\bf TS-2} and the second and third ones to {\bf TS-1}.
{\sc MTsearch} can find {\bf TS-1} and {\bf TS-2} also by starting from guess modes which are based on chemical intuition (see Supporting Information, Section 1).

The standard EVF optimization from hydroxymethylene following the lowest vibrational frequency mode does not converge to a TS, but
 falls back to the minimum-energy structure. Already a minor distortion of the minimum-energy structure towards the TS can already lead to
a successful location of {\bf TS-2} (see Supporting Information, Section 4.2, for details). 

Besides the TS optimization starting from trans-hydroxymethylene, 
we have also carried out TS localizations from formaldehyde and cis-hydroxymethylene. 
For cis-hydroxymethylene,  {\bf TS-2} was found by following the LST
guess mode. For formaldehyde, neither a LST guess mode nor a guess
mode based on chemical intuition led to convergence to {\bf TS-1}.

\subsection*{\sffamily \large Intramolecular proton-transfer reaction in a hydrazine Mo complex}

In this section, we study the hydrazine intermediate of Schrock's nitrogen-reducing catalyst and a generic model complex with aryl groups substituted by methyl groups. 
For the Schrock hydrazine complex, the lowest-energy spin state is a doublet. 
All spin states with higher multiplicity are at least 60 kJ/mol higher in energy.\cite{Schenk2009} 
We investigated the transition-state structure for a proton shift reaction from the nitrogen atom
 of N$_2$H$_4$ that ligates to molybdenum to one of the amido nitrogen atoms.

\begin{figure}[H]
\begin{center}
  \includegraphics[scale=1.0]{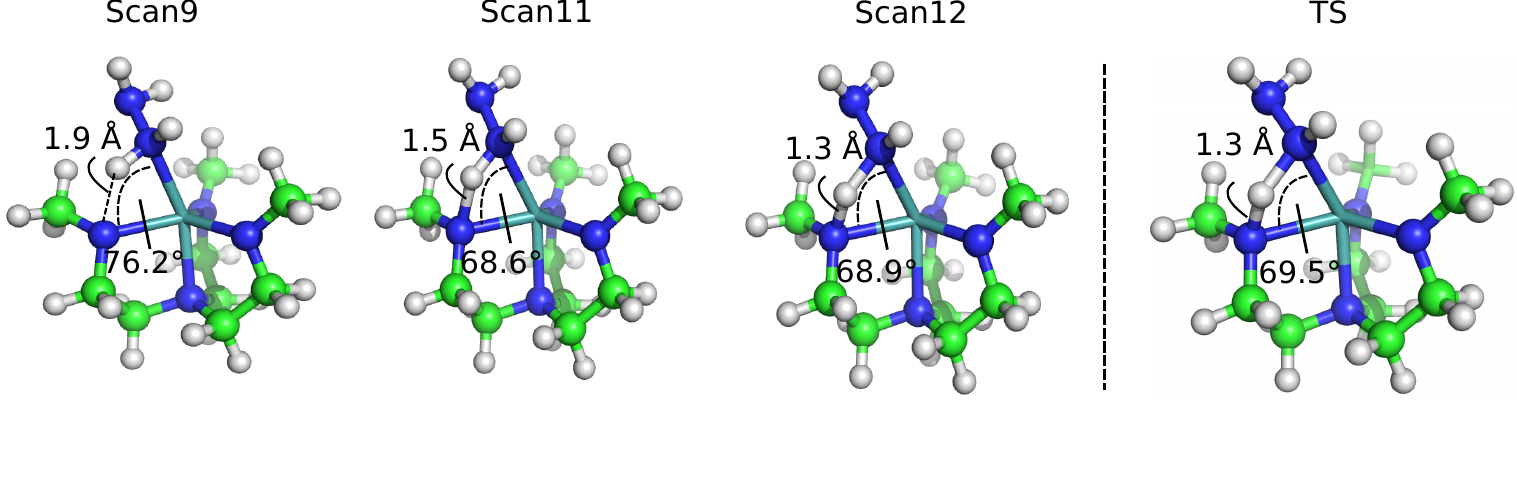} 
  \caption{Initial guess structures chosen from a constrained optimization scan along one H(N$_2$H$_4$)-N$_{\textrm{amido}}$ distance.
{\bf Scan9}, {\bf Scan11} and {\bf Scan12} are the 9th, 11th and 12th structure from a 12-step constrained optimization scan with the program {\sc Gaussian} (0.2 {\AA} increase of
the N$_{\mathrm{amido}}$-H distance in each step) 
along the N$_{\mathrm{amido}}$-H distance starting
from the hydrazine bound complex.}
\label{scan}
\end{center}
\end{figure}

For the generic model complex, we carried out a constrained optimization scan along the N$_{\mathrm{amido}}$-H distance that changes from reactant to product in 12 steps
including reactants and products of 0.2 {\AA}
step size to obtain a guess structure close to the TS.
From this constrained optimization scan, we selected the highest-energy structure ({\bf Scan12}, see Figure \ref{scan})
and an initial guess mode based on the structures {\bf Scan11} and the product. 
This mode has been simplified by retaining only those entries that refer to 
the transferring hydrogen atom.
This restriction to the ``moving'' part in the system improves convergence as other motions of parts of the system are discarded.
Moreover, it produces a guess mode that is transferable between homologous species (see the large complex below).

\begin{center}
[Figure 4 about here.]
\end{center}

The frequency analysis of structures {\bf Scan11} and {\bf Scan9} revealed that the lowest eigenvalue modes do not correspond to the desired transition vector. 
We observed that the standard EVF algorithm often fails to find the TS in this situation (see Supporting Information for details). Only for structure {\bf Scan12}, 
which is already very close to the TS, the standard EVF optimization converges
towards the TS. By contrast, {\sc MTsearch} was able to find the TS also from {\bf Scan11} and {\bf Scan9} 
(see Supporting Information, Section 4.3).

The root-mean-square deviation between the {\sc MTsearch}-optimized transition-state structures
and the one calculated with {\sc Turbomole}'s EVF is only 0.04 {\AA}, which means that the two algorithms converged to the same structure. 
The vibrational analysis revealed exactly one imaginary frequency of -i1244 cm$^{-1}$,
and the intrinsic reaction coordinates (IRC) connect the reactant and product structures, which confirms that we found the desired TS.
The stationary points calculated by {\sc MTsearch} are shown in Figure \ref{ts-schrock-small}.

It is noteworthy that the initial guess modes cannot only be obtained from a constrained scan, but also from a LST or NEB pathway or based on chemical intuition. 
For isomerization reactions, in which only one atom re-positions, as in our example, it is straightforward to manually choose an approximate transition pathway 
(cf.\ Supporting Information). However, the manual set-up of a proper molecular distortion that is likely to resemble a reaction pathway is possible 
and potentially useful also for other types of reactions.

In a Mode-Tracking-based TS search, one should confirm whether the very first mode converged with Mode-Tracking corresponds to the desired 
reaction pathway, since all following optimization steps are based on the direction of this initial mode. However, this can be done automatically
by calculating the overlap of the initial guess mode and the converged one. If the initial mode is not close to the transition vector, the optimization 
may lead to a different TS than the desired one (in our case, the TS for a rotation of the terminal NH$_2$ moiety of the N$_2$H$_4$ ligand was often found when the
initial guess mode was not clearly dominated by the shifting proton).

\begin{center}
 [Figure 5 about here.]
\end{center}
 
For the optimization of the analogous transition-state structure in the significantly larger hiptN$_3$N-ligated Schrock Mo catalyst, we 
rotated the coordinating N$_2$H$_4$ ligand in the minimum-energy structure such that an initial guess structure comparable to the TS of 
the generic model complex is obtained. Then, we performed a constrained 
optimization (fixed atoms are: molybdenum, the proton that moves and the two nitrogen atoms to which the proton binds in the reactant structures).
We choose as an initial guess mode the converged mode of the TS in the generic model complex (after alignment of the large and small homologous complexes, and choosing only
those entries for atoms that occur in both complexes; all other entries are set to zero).
Due to the significantly larger system size, the maximum number of orthogonal optimization steps performed within {\sc MTsearch}
is increased to 10. The stationary points with the TSs calculated by {\sc MTsearch} are displayed in Figure \ref{ts-schrock-small}.

With Mode-Tracking we obtain one imaginary frequency of -i1323 cm$^{-1}$ for the TS, 
which is similar to the one of the TS in the small model system. The mode is located on the proton that shifts.
To study the performance of our algorithm, we also calculated the full Hessian and obtained one negative frequency mode of -i1338 cm$^{-1}$.

Since the complete Hessian calculation and diagonalization for this molecule consisting of 284 atoms takes significantly longer than the Mode-Tracking calculation 
(about a week vs. 2 hours on 12 cores on a blade system featuring two six-core AMD Opteron 2435 processors (i.e., a total of 12 cores)), 
the computational time needed for the TS search is considerably reduced.

\section*{\sffamily \Large CONCLUSIONS}

The search for multiple reaction pathways starting from one minimum structure is still a main obstacle of current transition-state optimization programs.
In general, chemical intuition is needed to choose a suitable starting mode, which connects the reactants with the products.

In this paper, we presented an algorithm that efficiently combines the calculation of selected normal modes by the 
Mode-Tracking scheme\cite{Reiher2003} and the eigenvector-following procedure to locate and optimize transition-state structures.
Since Mode-Tracking avoids the time-consuming calculation of the complete Hessian matrix and instead only optimizes the modes of interest,
{\sc MTsearch} is particularly suitable for optimizing transition-state structures of large reactive molecular systems. 
The search for several transition-state structures is feasible and the starting structures for a search
may lie outside the quadratic region of a transition-state structure.

We investigated our algorithm at the example of four intramolecular reaction pathways: the rotational barrier of C$_2$H$_6$, the isomerization of H$_2$CO,
a proton shift in the hydrazine-bound intermediate of the [Mo\-(hipt\-N$_3$N)] catalyst by Schrock as well as in a model system 
with methyl substituents instead of the hipt substituents.
Initial guess modes for the Mode-Tracking procedure can be extracted either from an LST or NEB pathway or based on chemical intuition.
A TS optimization can be started either from two or from only one minimum-energy structure.
The potential energy surface can be explored in a customized way along the desired directions.
Even for a large molecule such as the hydrazine-coordinating Mo complex of Schrock and coworkers with more than 200 atoms, we were able to efficiently locate a transition-state structure.

By choosing different initial guess modes and/or branching off at certain structures during the
optimization, one may scan the potential energy surfaces along different directions simultaneously.

\subsection*{\sffamily \large ACKNOWLEDGMENTS}
This work has been financially supported by ETH Zurich and the Schweizer Nationalfonds (SNF project 200021L\_156598).
MB gratefully acknowledges support by a fellowship of the Fonds der Chemischen Industrie (FCI).
CH acknowledges funding by the FCI.

\clearpage




\clearpage

\clearpage

\begin{figure}[H]
\begin{center}
 \includegraphics[scale=0.7]{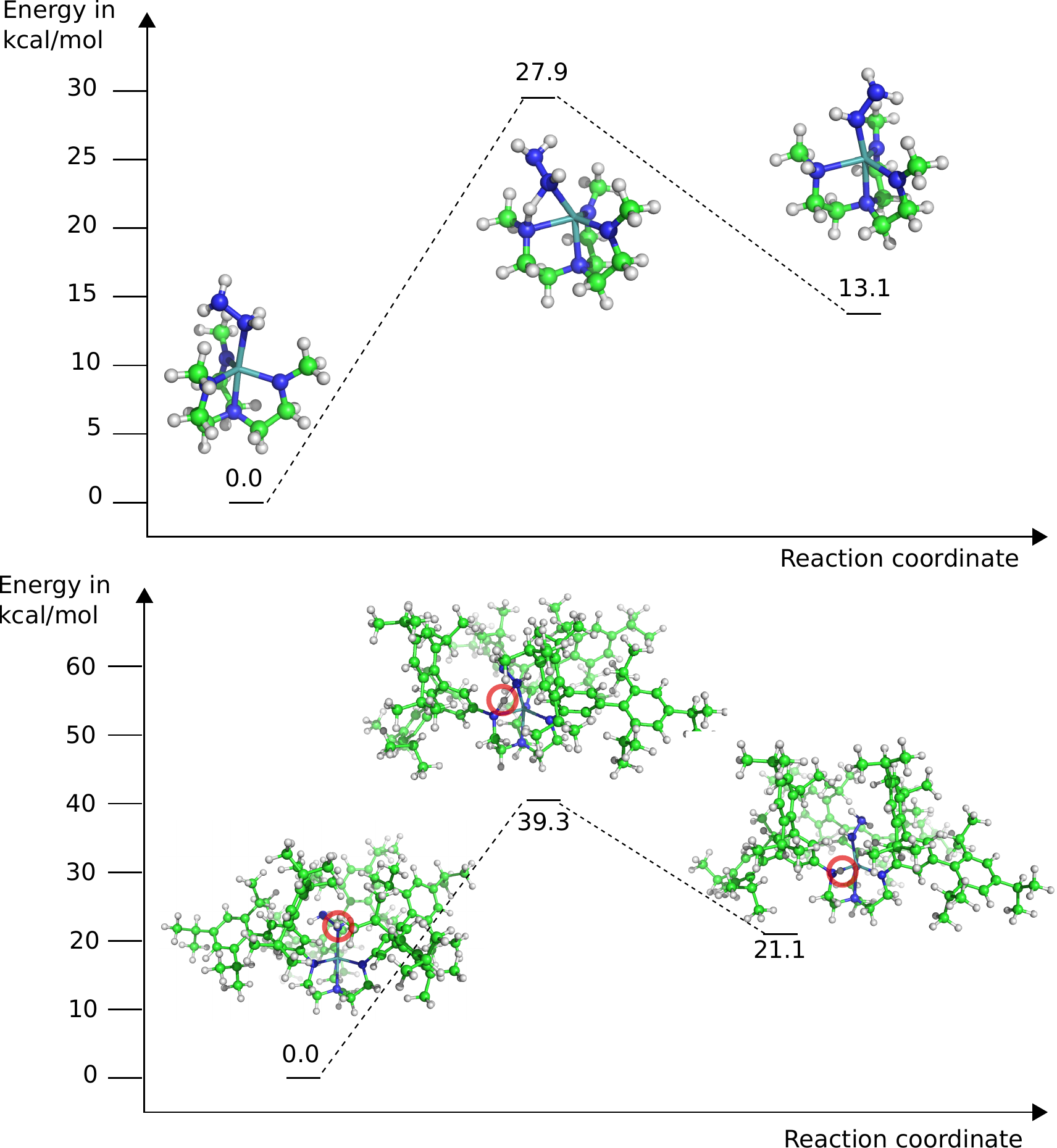}
\caption{\label{ts-schrock-small} Internal proton shift reaction pathway of the generic Schrock model system (top) and the full Schrock hydrazine Mo complex (bottom) 
with BP86/RI/def2-SV(P) transition-state structures optimized with {\sc MTsearch}.
Element color code: green, C; blue, N; cyan, Mo; white, H. The red circles highlight the proton that shifts during the reaction.}
\label{ts-schrock-small}
\end{center}
\end{figure}

\end{document}